\documentclass[sigconf]{acmart}

\usepackage{booktabs} 
\usepackage{amsmath,amssymb,amsfonts}
\usepackage{algorithmic}
\usepackage[linesnumbered,ruled]{algorithm2e}
\usepackage{graphicx}
\usepackage{textcomp}
\usepackage{xcolor}
\usepackage{multirow}
\usepackage{bbm}
\usepackage{xcolor,colortbl}
\definecolor{green}{rgb}{0.5,0.9,0.0}
\newcommand{\optimalthreeglatency}{\cellcolor{green}3.1}  
\newcommand{\optimalthreegenergy}{\cellcolor{green}6.6}  
\newcommand{\optimalfourglatency}{\cellcolor{green}1.8}  
\newcommand{\optimalfourgenergy}{\cellcolor{green}4.1}  
\newcommand{\optimalwifilatency}{\cellcolor{green}1.6}  
\newcommand{\optimalwifienergy}{\cellcolor{green}3.5}  

\def\BibTeX{{\rm B\kern-.05em{\sc i\kern-.025em b}\kern-.08em
    T\kern-.1667em\lower.7ex\hbox{E}\kern-.125emX}}

\renewcommand\footnotetextcopyrightpermission[1]{} 
\begin{document}
\title{BottleNet: A Deep Learning Architecture for Intelligent Mobile Cloud Computing Services}

\begin{abstract}

Recent studies have shown the latency and energy consumption of deep neural networks can be significantly improved by splitting the network between the mobile device and cloud. This paper introduces a new deep learning architecture, called BottleNet, for reducing the feature size needed to be sent to the cloud. Furthermore, we propose a training method for compensating for the potential accuracy loss due to the lossy compression of features before transmitting them to the cloud. BottleNet achieves on average 30$\times$ improvement in end-to-end latency and 40$\times$ improvement in mobile energy consumption compared to the cloud-only approach with negligible accuracy loss.

\end{abstract}

\keywords{deep learning, collaborative intelligence, mobile computing, cloud computing, feature compression}

\author{Amir Erfan Eshratifar}
\affiliation{
\department{Department of Electrical Engineering}
\institution{University of Southern California}
\city{Los Angeles}
\state{California}
}
\email{eshratif@usc.edu}

\author{Amirhossein Esmaili}
\affiliation{
\department{Department of Electrical Engineering}
\institution{University of Southern California}
\city{Los Angeles}
\state{California}
}
\email{esmailid@usc.edu}

\author{Massoud Pedram}
\affiliation{
\department{Department of Electrical Engineering}
\institution{University of Southern California}
\city{Los Angeles}
\state{California}
}
\email{pedram@usc.edu}

\maketitle

\section{Introduction}

Mobile and Internet of Things (IoT) devices are increasingly relying on deep neural networks (DNNs) to provide state-of-the-art performance in various intelligent applications \cite{krizhevsky2012imagenet, Resnet, girshick2016region, hinton2012deep, mikolov2013distributed}. Due to limited computational and storage resources of mobile devices, which prohibits full deployment of advanced deep models on these devices (the \textit{mobile-only} method), the most common deployment approach of most of the DNN-based applications on mobile devices relies on using the cloud. In this approach, which is referred to as the \textit{cloud-only} approach, the deep network is fully placed on the cloud, and thus the input is sent from the mobile to cloud for performing the computations associated with the inference network, and the output is sent back to the mobile device. 

The cloud-only approach requires mobile devices to send vast amounts of data (e.g. images, audios and videos) over the wireless network to the cloud. This can give rise to considerable energy and latency overheads on the mobile device. Furthermore, pushing all computations toward the cloud can lead to congestion in a scenario where a large number of mobile devices simultaneously send data to the cloud. As a compromise between the mobile-only and the cloud-only approach, recently, a body of research work has been investigating the idea of splitting a deep inference network between the mobile and cloud \cite{eshratifar2018jointdnn,EshratifarGLS, kang2017neurosurgeon,grulich2018collaborative,chen2018intermediate,choi2018near,choi2018deep}. In this approach, which is referred to as \textit{collaborative intelligence}, the computations associated with initial layers of the inference network are performed on the mobile device, and the feature tensor (activations) of the last computed layer is sent to the cloud for the remainder of computations. The main motivation for collaborative intelligence is the fact that in many applications which are based on convolutional neural networks (CNNs), the feature volume of layers will shrink in size as we go deeper in the model \cite{kang2017neurosurgeon,eshratifar2018jointdnn,choi2018deep}. Therefore, computing a few layers on the mobile and then sending the last computed feature tensor to the cloud can reduce the latency and energy overheads of wireless transfer of the data to the cloud compared to sending the input of the model directly to the cloud. Furthermore, pushing a portion of computations onto the mobile devices can reduce the congestion on the cloud and hence increase its throughput.

\begin{figure}
\begin{center}
  \includegraphics{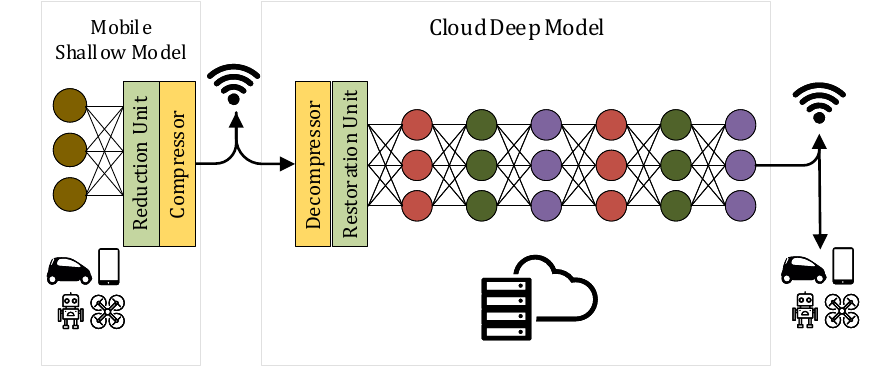}
  \caption{Overview of the proposed method.} 
  \label{fig:overall_architecture}
\end{center}
\end{figure}

In research studies investigating collaborative intelligence, a given deep network is split between the mobile device and the cloud without any modification to the network architecture itself \cite{eshratifar2018jointdnn,kang2017neurosurgeon,grulich2018collaborative,chen2018intermediate,choi2018near,choi2018deep}. In this paper, we investigate altering the underlying deep model architecture to make it collaborative intelligence friendly. For this purpose, we mainly focus on altering the underlying deep model in a way that the feature data size needed to be transmitted to the cloud is reduced. This is because in the studies investigating collaborative intelligence, the latency and energy overheads of the wireless data transfer to the cloud yet play a major role in the total mobile energy consumption and the end-to-end latency \cite{eshratifar2018jointdnn}. Therefore, reducing the transmitted data size to cloud is generally beneficial. For this purpose, we add a non-expensive learnable \textit{reduction unit} after the layers assigned to be computed on the mobile device, and the output of this unit is then compressed using conventional compressors (e.g., JPEG) and sent to the cloud. Correspondingly, a decompressor and a learnable \textit{restoration unit} is added before the layers assigned on the cloud. The main components of reduction and restoration units are convolutional layers which their dimensions are determined in a way that the input of the reduction unit and the output of the restoration unit have the same dimensionality. 

An overview of the proposed method is shown in Fig.~\ref{fig:overall_architecture}. The insertion location and size of the the reduction and restoration units in the underlying DNN are determined as explained in Section \ref{section.profiling}. Since by inserting the reduction unit, a data bottleneck is created in the model, the combination of the learnable reduction unit, compressor and decompressor, and the learnable restoration unit is referred to as the \textit{bottleneck unit}, and the new network architecture including the bottleneck unit is referred to as \textit{BottleNet}, which is trained end-to-end. For the reduction unit, we evaluate and compare dimension reductions along both the channel and spatial dimensions of intermediate feature tensors as explained in Section \ref{section.bottleneck_unit}. 

As we see in Section \ref{section.evaluation}, an obvious benefit of using the proposed bottleneck unit is in deep models where feature tensor sizes are relatively high, such as ResNet \textcolor{black}{\cite{Resnet}}. In such networks, the layer in which the feature size is less than the input size is either not present or lies very deep inside in the network. Therefore, if we want to merely split the network and send the intermediate feature tensor to the cloud as in  \cite{kang2017neurosurgeon,eshratifar2018jointdnn}, we need to compute considerable number of  layers on the mobile. This will push a major workload to the mobile which will likely result in higher latency and energy consumption compared to the cloud-only approach. This could be the main reason that previous work on collaborative intelligence usually has focused on deep architectures where the intermediate feature size is relatively small compared to the input size after computing only a few layers, such as AlexNet \cite{krizhevsky2012imagenet} and DeepFace \cite{taigman2014deepface}. 

Furthermore, since features of an intermediate layer in a deep model tend to exhibit statistical characteristics such as data redundancy and lower entropy compared to the input of the model, compressing the feature tensor before sending it to the cloud can potentially achieve considerable reductions in the data size needed to be sent over the wireless network. Therefore a major portion of the works studying collaborative intelligence also consider feature compression instead of direct transfer of the feature tensor to the cloud \cite{eshratifar2018jointdnn,chen2018intermediate,choi2018near,choi2018deep, isqed2019}. In this work, we consider lossy compression of the feature tensor. Lossy compression methods can lead to higher bit savings compared to the lossless compression approaches. However, they may adversely affect the achieved accuracy and thus lossy compression methods are less studied in the works using feature compression in collaborative intelligence framework, as they mostly use lossless compression techniques on the features before transmitting them to the cloud. In order to compensate for the reduced accuracy due to the lossy compressor, we propose a novel training method for the network which approximates the behavior of the lossy compressor as an identity function in backpropagation. The proposed training method is explained in detail in Section \ref{section.compression-aware_training}.         

In summary, the contributions of this paper are as follows:
\begin{itemize}
    \item We propose the bottleneck unit, in which by using a learnable reduction unit followed by a lossy compressor, the feature tensor size required to be transmitted to the cloud is significantly reduced.
    \item For training our model, we propose a lossy compression-aware training method in order to compensate for the accuracy loss.
    \item Using the proposed BottleNet architecture, we achieve on average 30$\times$ improvement in end-to-end latency and 40$\times$ improvement in mobile energy consumption compared to the cloud-only approach with negligible accuracy loss.   
\end{itemize}

The remainder of the paper is structured as follows: Section \ref{section.proposed_method} provides a more detailed explanation of the bottleneck unit and training method. Section \ref{section.evaluation} provides the energy and latency improvements of the proposed bottleneck unit and discusses the efficacy of our training approach, and also elaborates on the flexibility of changing the partition point depending on the cloud server congestion and wireless network conditions. Finally, Section \ref{section.conclusion} concludes the paper. 

\begin{figure}
\begin{center}
  \includegraphics[width=\columnwidth]{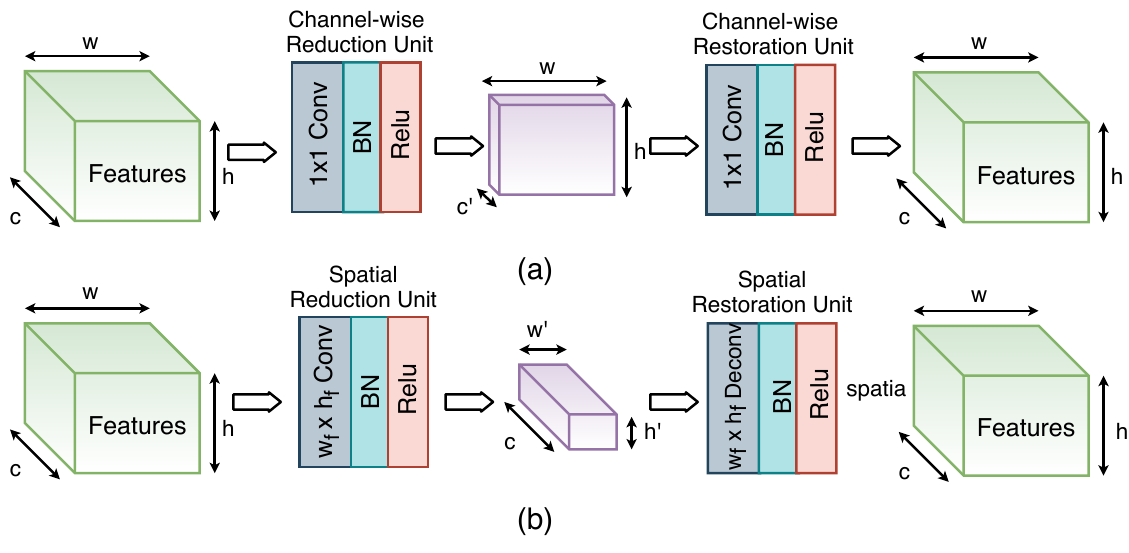}
  \caption{Learnable dimension reduction and restoration units along the (a) channel and (b) spatial dimension of features. }
  \label{fig:spatial_channel_reduction}
\end{center}
\end{figure}

\section{Proposed Method} \label{section.proposed_method}
In this section, first, we describe details of the bottleneck unit. Then, we explain our proposed training method when a non-differentiable lossy compression is applied to the intermediate feature tensor before transmitting it to the cloud. Finally, we explain our approach for finding the best insertion location \textcolor{black}{and size} of the bottleneck unit in the underlying deep model to achieve the lowest end-to-end latency and/or mobile energy consumption in different wireless settings.

\begin{figure}
\begin{center}
  \includegraphics[width=\columnwidth]{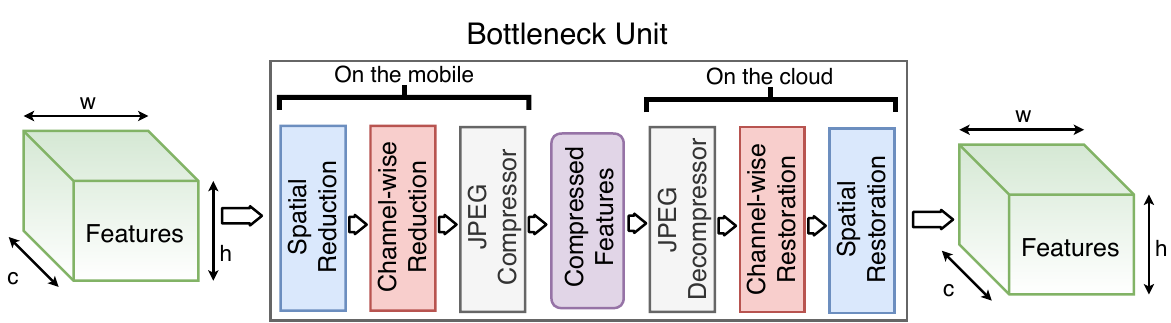}
  \caption{The bottleneck unit embedded with a non-differentiable lossy compression (e.g., JPEG).}
  \label{fig:bottleneck}
\end{center}
\end{figure}

\subsection{Bottleneck Unit} \label{section.bottleneck_unit}
For dimension reduction in the feature tensor in the bottleneck unit, we evaluate and compare dimension reductions along either channel or spatial dimensions. The bottleneck unit, referred to as autoencoder in deep learning context, is responsible for learning a dense representation of the features in an intermediate layer. As depicted in Fig.~\ref{fig:spatial_channel_reduction}, channel-wise reduction, shrinks the number of channels of the features, and spatial reduction shrinks the spatial dimensions (width and height) of the features. More specifically, channel-wise reduction unit takes a tensor of size ($batch\_size$, $w$, $h$, $c$) as input, and outputs a tensor of size ($batch\_size$, $w$, $h$, $c'$) by applying a convolution filter of size ($1$, $1$, $c$, $c'$) followed by normalization and non-linearity layers. The output tensor of the reduction unit is the reduced-order representation of its input ($c'$ $\ll$ $c$). Spatial reduction unit takes a tensor of size ($batch\_size$, $w$, $h$, $c$) as input, and outputs a tensor of size ($batch\_size$, $w'$, $h'$, $c$) by applying a convolution filter of size ($w_f$, $h_f$, $c$, $c$) followed by normalization and non-linearity layers. The output tensor of the reduction unit is the reduced-order representation of its input ($w'< w$, and $h' < h$). In both channel-wise and spatial reduction units, we use ReLU as the non-linearity function. For reduction in the spatial dimension, the stride step of the convolution should be more than one. It should be noted that to cover each neuron during the convolution at least once, the size of this filter should be more than the stride step size, i.e., $w_f > \frac{w}{w'}$, and $h_f > \frac{h}{h'}$. In this paper, we use the same reduction factor for both width and height, referred to as the spatial reduction factor of $s$, i.e., $\frac{w}{w'}=\frac{h}{h'}=s$. 

The bottleneck unit architecture uses both spatial and channel-wise reduction units followed by a compressor unit on the mobile device to create a compressed representation of the feature tensor which is the tensor transmitted to the cloud. On the cloud, the bottleneck unit uses a decompressor followed by channel-wise and spatial restoration units to restore the dimension of the feature tensor. The detailed architecture of the bottleneck unit is depicted in Fig. \ref{fig:bottleneck}. The choice of ReLU in reduction units can potentially lead to higher zero rates resulting in higher compression ratios. The bottleneck unit is inserted between two selected consecutive layers of the underlying deep model, where these two layers are selected by an algorithm as explained in Section \ref{section.profiling}.

\subsection{Non-differentiable Lossy Compression Aware Training} \label{section.compression-aware_training}
Lossy compression methods result in higher bit savings compared to lossless methods. However, lossy compression is inherently a non-differentiable function. Specifically, quantization
is an integral part of the compression and is not differentiable. Introducing non-differentiable functions in a neural network disables the back-propagation because the gradients are not propagated to the layers before the non-differentiable function, resulting in the model not end-to-end trainable. To solve this issue, we introduce a new training method to enable the model to be end-to-end differentiable by defining a gradient for the pair of compressor and decompressor during the backpropagation. In other words, the pair of lossy compressor and decompressor is used as is during the forward propagation, while we treat this pair as an identity function during backpropagation (i.e., gradients passed without any change to the layers before the compressor). 
Therefore, the whole model will become end-to-end differentiable. The effectiveness of using this training method instead of simply training of the model without considering the compression will be explained in Section \ref{section.efficacy_training}.

The input to the compressors are typically quantized to unsigned n-bit numbers by an uniform quantizer. Similar to \cite{choi2018near}, to quantize features, $F$, we use following quantizer: 
\begin{equation}
\Tilde{F} = round (\frac{F-min(F)}{max(F)-min(F)} * (2^n - 1))
\end{equation}
In addition, the input to the compressors should be reshaped into 2-D tensors. The features, $F$, with $C$ channels, are rearranged in a tiled image with the width of $2^{ceil(\frac{1}{2} log_2(C))}$ and the height of $2^{floor(\frac{1}{2} log_2(C))}$ to keep the aspect ratio as square as possible to achieve the maximum compression ratio.

\begin{figure}
\begin{center}
  \includegraphics[width=\columnwidth]{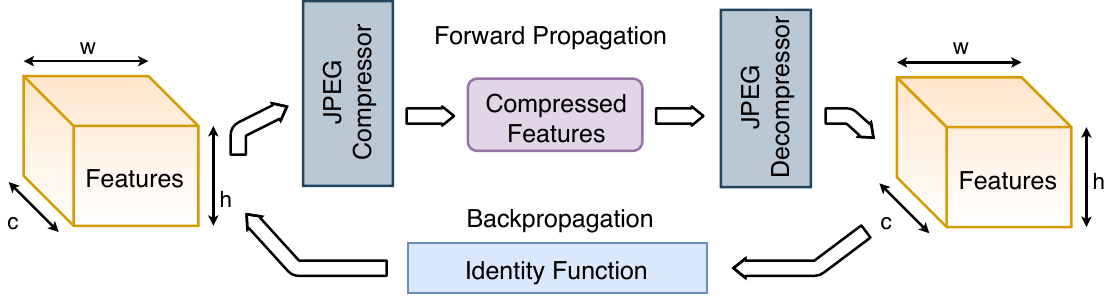}
  \caption{Embedding non-differentiable compression (e.g., JPEG) in DNN architecture. We approximate the pair of JPEG compressor and decompressor units by identity function to make the model differentiable in backpropagation.}
  \label{fig:dnn_jpeg}
\end{center}
\end{figure}

\subsection{Architecture Selection} \label{section.profiling}

The proposed algorithm, for choosing the location of the bottleneck unit and the proper value of $c'$ (reflecting the degree of reduction along the channel dimension), and \textcolor{black}{$s$} (reflecting the degree of reduction along the spatial dimension) comprises of three main steps: 1) Training, 2) Profiling, and 3) Selection.  We consider placing the bottleneck unit each time after an arbitrary selected layer of the underlying network, for total of $M$ different locations in the network, where $M$ is less than or equal to total $N$ layers of the network. In each of $M$ locations, we train different architectures associated with different degrees of dimension reduction along channel or spatial dimensions, and among those which result in acceptable accuracy levels, we select the one with the minimum bit requirement. We repeat this process for all of $M$ locations. At the end, among $M$ selected models associated with $M$ different partitioning solutions of the network, depending on our optimization target, we choose the best partitioning in terms of minimizing mobile energy consumption and/or end-to-end latency. We measure the latency and mobile energy consumption of computations assigned to the mobile (including reduction and compressor units), wireless transfer of dense compressed feature tensor to the cloud, and computations assigned to the cloud (including decompressor and restoration units). 

The detailed algorithm for choosing the location of the bottleneck unit and the proper amount of reductions in channel and spatial dimensions is presented in Algorithm~\ref{the_algorithm}.   

\begin{algorithm}
    \SetKwInOut{Input}{Input}
    \SetKwInOut{Output}{Output}

\textbf{Inputs:}\\
$N$: number of layers in the DNN\\
$M$: number of partitioning points in the DNN ($M \leq N$)\\
$bottleneck(s,c')$: A bottleneck unit with the spatial reduction factor of $s$ and reduced channel size of $c'$ \\
$S_{max}$: maximum allowable spatial reduction factor\\
$C'_{max}$: maximum allowable reduced channel size\\
$K_{mobile}$: current load level of mobile\\
$K_{cloud}$: current load level of cloud\\
$t_{mobile}, p_{mobile}(j,K_{mobile}) |j=1..M$: latency and power on the mobile corresponding to partition $j$ and load $K_{mobile}$\\
$t_{cloud}(j,K_{cloud}) |j=1..M$: latency on the cloud corresponding to partition $j$ and load $K_{cloud}$\\
$NB$: wireless network bandwidth\\
$PU$: wireless network up-link power consumption\\
\textbf{Outputs:}\\
Best partitioned model\\
\textbf{Variables: }\\
$\{D_j|j=1..M\}$: compressed feature size in each of $M$ partitioning solutions\\
~\\
// Training phase \\
 \For{$j = 1;\ j \leq M;\ j = j + 1$}{
    \For{$c' = 1;\ c' \leq C'_{max};\ c' = c' + 1$} {
        \For{$s = 1;\ s \leq S_{max};\ s = s + 1$} {
            Place $bottleneck(s, c')$ after j-th layer\\
            Train()\\
            Store the corresponding model and its accuracy
        }
    }
 }
 
  \For{$j = 1;\ j \leq M;\ j = j + 1$}{
    For those models that the bottleneck unit is placed after $j$-th layer, among ones with acceptable accuracy, store the one with minimum compressed feature size and store its compressed feature size as $D_j$\\
  }
 ~\\
 // Profiling phase \\
  \For{$j = 1;\ j \leq M;\ j = j + 1$}{
    $TM_j$ = $t_{mobile}(j, K_{mobile})$\\
    $PM_j$ = $p_{mobile}(j, K_{mobile})$\\
    $TC_j$ = $t_{cloud}(j, K_{cloud})$\\
    $TU_j$ = $D_{j}/NB$
    }
    ~\\
// Selection phase \\
     \If{target is min latency} {
    	return $argmin_{j=1..M} (TM_j + TU_j + TC_j)$
      }
      \If{target is min energy} {
        return $argmin_{j=1..M} (TM_j \times PM_j + TU_j \times PU)$
      }
    \caption{The partitioning algorithm for BottleNet}
    \label{the_algorithm}
\end{algorithm}

\section{Evaluation} \label{section.evaluation}


\begin{figure*}
  \begin{center}
  \includegraphics[width=\textwidth]{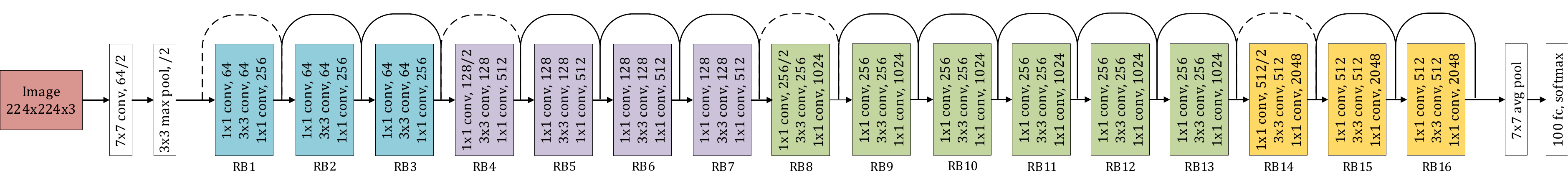}
  \caption{ResNet-50 architecture and its 16 residual blocks.}
  \label{fig:resnet50}
  \end{center}
\end{figure*}

\begin{figure}
\begin{center}
  \includegraphics[width=\columnwidth]{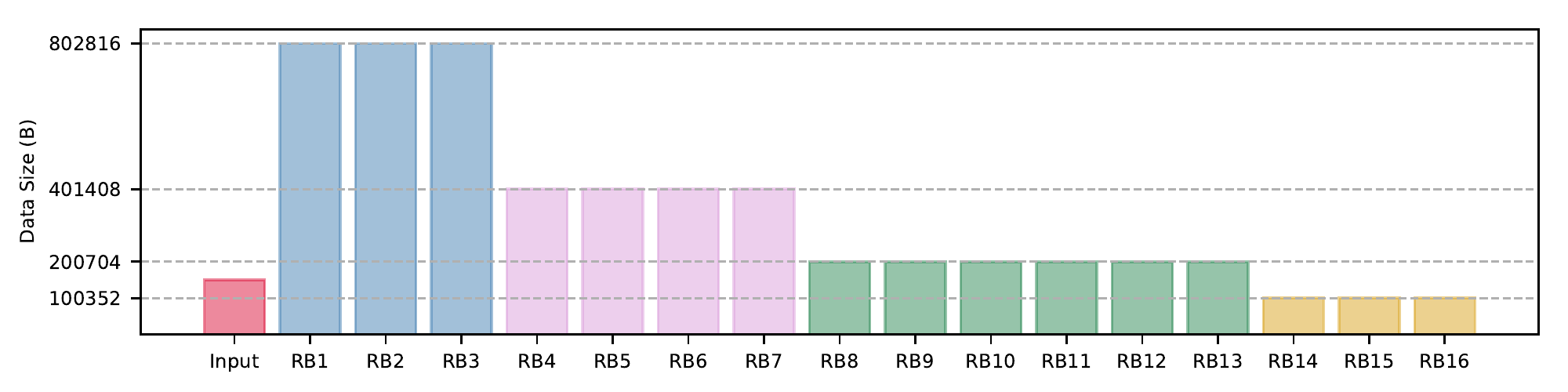}
  \caption{Input image size of the model and the size of output feature tensor of each residual block in ResNet-50.}
  \label{fig:resnet_feature_size}
\end{center}
\end{figure}

\subsection{Experimental Setup}\label{AA}
We evaluate our proposed method on NVIDIA Jetson TX2 board~\cite{JetsonTX2} equipped with NVIDIA Pascal\texttrademark\space GPU with 256 CUDA cores which fairly represents the communication and computation resources of mobile devices. Our server platform is equipped with a NVIDIA Geforce\textregistered\space GTX 1080 Ti GPU, which has almost 30x more computing power compared to our mobile platform. The detailed specifications of our mobile and server platforms are presented in Table \ref{table:mobile_platform} and Table \ref{table:server_platform}, respectively. We measure the GPU power consumption on our mobile platform using INA226 power monitoring sensor with sampling rate of 500~KHz~\cite{INA226}. 
For the wireless network settings, the average upload speed of different wireless networks, 3G, 4G, and Wi-Fi, in the U.S. are used in our experiments~\cite{MobNet,Speedtest}. We use the transmission power models of \cite{4GLTE} for wireless networks with estimation error rate of less than 6\%. The power level for up-link is estimated by $P_u = \alpha_u t_u + \beta$ 
, where $t_u$ is the up-link throughput, and $\alpha_u$ and $\beta$ are regression coefficients of power models. The values for our power model parameters are presented in Table~\ref{table:network_parameters}.

We prototype the proposed method by implementing the inference networks for both the mobile device and cloud server using NVIDIA TensorRT\texttrademark\space\cite{TensorRT}, which is the state-of-the-art platform for high-performance deep learning inference. It includes a deep learning inference optimizer and run-time that delivers low latency and high-throughput for deep learning inference applications. TensorRT is equipped with cuDNN\cite{cuDNN}, a GPU-accelerated library of primitives for deep neural networks. TensorRT supports three precision modes for creating the inference graph, namely FP32 (single precision), FP16 (half precision), and INT8 (8-bit integer). However, our mobile device does not support INT8 operations on its GPU for inference. Therefore, we use FP16 mode for creating the inference graph from the trained model graph, where for the training itself single precision mode is used. As demonstrated in \cite{deepcompression}, 8-bit quantization would be enough for even challenging tasks like ImageNet~\cite{ImageNet} classification. Therefore, we apply 8-bit quantization on FP16 data types, using the uniform quantizer presented in Section~\ref{section.compression-aware_training}, before applying the lossy compression on them. We implement our client-server interface using Thrift~\cite{slee2007thrift}, an open source flexible RPC interface for inter-process communication. Given a partition decision, execution begins on the mobile device and cascades through the layers
of the DNN leading up to that partition point. Upon completion of that layer and the reduction unit and lossy compressor, mobile sends the reduced dense feature tensor from the mobile device to the cloud. Cloud server then executes the computations associated with the decompressor, restoration unit, and the remaining DNN layers. Upon the completion of the the execution of last DNN layer on the cloud, the inference result is sent back to the mobile device. For the choice of our lossy compressor, here, we use JPEG compression.

For evaluating our proposed method, we use ResNet-50 as our underlying deep model, which is one of the widely used deep models allowing training very deep networks by using skip connections in residual blocks (RBs).
 ResNet architecture comes with flexible number of layers such as 34, 50, 101. There are 16 RBs in ResNet-50. Using algorithm \ref{the_algorithm} explained in Section \ref{section.profiling}, we obtain 16 different models where each model is associated with placing the bottleneck unit after one of the 16 available RBs. As presented in algorithm \ref{the_algorithm}, the obtained $c'$ and $s$ of the bottleneck unit, when it is placed after each of 16 RBs, corresponding to 16 different partitions, could be different, where $c'$ and $s$ were corresponding to different degrees of reductions in channel and spatial dimensions, respectively. 
 The architecture of ResNet-50 is presented in Fig.~\ref{fig:resnet50}. The input image size of the model and the size of output feature tensor of each RB are presented in Fig.~\ref{fig:resnet_feature_size}.
As indicated in Fig.~\ref{fig:resnet_feature_size}, the size of intermediate feature tensors in ResNet-50 are larger than the input size up until RB14, which is relatively deep in the model. Therefore, merely splitting this network between the mobile and cloud for collaborative intelligence may not perform better than the cloud-only approach in terms of latency and mobile energy consumption, since a large portion of the workload is pushed toward the mobile.

For our dataset, we use miniImageNet~\cite{miniImageNet}, a subset of ImageNet, which includes 100 classes and 600 examples per each class. 85\% of whole dataset examples are used as the training set, and the rest as the test set. We randomly crop a 224$\times$224 region from each sample for data augmentation. For training of our models, we use 90 epochs of data.  

\begin{table}[ht]
\caption{Mobile device specifications} 
\centering 
\begin{tabular}{|c|c|} 
\hline 
\textbf{Component}& \textbf{Specification} \\ [0.5ex] 
\hline 
System & NVIDIA Jetson TX2 Developer Kit \\
\hline
GPU & NVIDIA Pascal\texttrademark, 256 CUDA cores \\
\hline
CPU & HMP Dual Denver + Quad ARM\textregistered\space A57/2 MB L2 \\
\hline
Memory & 8 GB 128 bit LPDDR4 59.7 GB/s \\ 
\hline 
\end{tabular}
\label{table:mobile_platform} 
\end{table}

\begin{table}[ht]
\caption{Server platform specifications} 
\centering 
\begin{tabular}{|c|c|} 
\hline 
\textbf{Component}& \textbf{Specification} \\ [0.5ex] 
\hline 
GPU & NVIDIA Geforce\textregistered\space GTX 1080 Ti, 12GB GDDR5\\
\hline
CPU & Intel\textregistered\space Xeon\textregistered\space CPU E7- 8837  @ 2.67GHz \\
\hline
Memory & 64 GB DDR4\\ 
\hline 
\end{tabular}
\label{table:server_platform} 
\end{table}

\begin{table}[h]
	\caption{\textcolor{black}{Wireless networks parameters}} 
	\centering 
	\begin{tabular}{|c|c|c|c|} \hline
	\textbf{Param.} & \textbf{3G} & \textbf{4G} & \textbf{Wi-Fi} \\ \hline
	$t_u$ (Mbps)	&	1.1		&	5.85	&	18.88 \\ \hline
$\alpha_u$ (mW/Mbps)	& 868.98	& 438.39	&	283.17    \\ \hline
$\beta$ (mW)	& 817.88	& 1288.04	&	132.86            \\ \hline
\end{tabular}
\label{table:network_parameters} 
\end{table}

\subsection{Latency and Energy Improvements}
The accuracy of ResNet-50 model for miniImageNet dataset without the bottleneck unit is 76\%, which we refer to as the target accuracy. By assuming an acceptable accuracy loss of 2\% compared to the target accuracy, using algorithm \ref{the_algorithm}, placing the bottleneck unit after RBs 1-3, 4-7, 8-13, and 14-16, requires the reduced channel size $c'$ of 1, 2, 5, and 10, respectively (placing the bottleneck unit after different RBs corresponds to different partitions). For all 16 partitions, the spatial factor reduction of $s$ is obtained as 2 by algorithm \ref{the_algorithm}. According to our experiments, defining the acceptable accuracy loss of 2\% allows channel-wise and spatial reduction units achieve significant dimension reductions and thus bit savings. For instance, when the bottleneck unit is placed after RB1, feature tensor of size ($56$, $56$, $256$) is reduced to ($28$, $28$, $1$) using the channel-wise and spatial reduction units. 

\begin{table*}[ht]
	\caption{The end-to-end Latency, mobile energy consumption, and offloaded data size for different partition points in ResNet-50 using the proposed method 
	} 
	\centering 
	\begin{tabular}{|c|c|c|c|c|c|c|c|c|c|c|c|c|c|c|c|c|} \hline
	\textbf{Layer} & RB1 &  RB2 &  RB3 &  RB4 &  RB5 &  RB6 &  RB7 &  RB8 &  RB9 &  RB10 &  RB11 &  RB12 &  RB13 &  RB14 &  RB15 & RB16 \\
	\hline
	\textbf{Offloaded Data (B)} & 316 &  317 &  314 &  166 &  171 &  168 &  170 &  96 &  90 &  98 & 101 & 101 & 95 &  52 & 52 &  53 \\
	\hline
	\textbf{Latency 3G (ms)} & \optimalthreeglatency & 4.1 &	4.9 &	5.2 &	6.3 &	7.5 &	8.2 &	9.6 &	10.7 &	11.6 &	12.8 & 13.4 & 14.8 &	15.1 &	16.0 &	17.1 \\
	\hline
	\textbf{Energy 3G (mJ)} & \optimalthreegenergy & 7.6 &	8.1 &	9.7 &	10.8 &	11.9 &	12.6 &	13.9 &	14.1 &	15.8 &	16.1 & 17.6 &	18.5 &	19.8 &	20.7 &	21.9\\
	\hline
	\textbf{Latency 4G (ms)} & \optimalfourglatency & 2.5 & 3.3 & 4.2 &	5.0 & 5.9 & 6.9 & 8.6 &	9.4 & 10.3 &	11.9 &	12.7 &	14.1 &	15.0 & 15.7 & 16.9\\
	\hline
	\textbf{Energy 4G (mJ)} & \optimalfourgenergy &	6.8 &	7.0 &	8.9 &	10.6 &	11.3 &	12.9 &	13.1 &	14.0 &	15.6 & 16.0 &	17.1 &	18.3 &	19.1 &	20.3 &	21.2\\
	\hline
	\textbf{Latency Wi-Fi (ms)} & \optimalwifilatency & 2.4 & 3.0 & 4.1 &	4.9 & 5.8 & 6.8 & 8.5 &	9.3 & 10.1 &	11.8 &	12.6 &	14.0 &	14.9 & 15.7 & 16.9\\
	\hline
	\textbf{Energy Wi-Fi (mJ)} & \optimalwifienergy &	5.6 &	6.1 &	7.4 &	9.5 &	10.8 &	12.3 &	12.5 &	13.8 &	14.9 & 15.6 &	16.9 &	18.1 &	19.0 &	20.1 &	21.0\\
	\hline
\end{tabular}
\label{table:layer_split} 
\end{table*}

\begin{table*}[ht]
\caption{Comparison of the proposed method with mobile-only and cloud-only approaches}
  \centering
\label{tab:lateny_energy_results}
\begin{tabular}{|c|c|c|c|c|c|c|c|}
\hline
\multicolumn{2}{|c|}{Setup}   & Latency (ms) & Energy (mJ) & Bottleneck Unit Location & Offloaded Data (B) & Accuracy\\
\hline
\multirow{1}{*}{Mobile-only}  & -   &   15.7    & 20.5 & - & 0  & 76.1 \\
\hline
\multirow{3}{*}{Cloud-only}  & 3G   &   196.2 & 310.1 & - & 26766 & 76.1 \\
                         & 4G  &   37.9    & 168.3 & - & 26766 & 76.1 \\
                          & Wi-Fi  &   13.1   & 110.7  & - & 26766 & 76.1 \\
\hline
\multirow{3}{*}{BottleNet}  & 3G   &   3.1    & 6.6 & After RB1 & 316  & 74.1\\
                      & 4G  &  1.8    & 4.1 & After RB1 & 316 & 74.1\\
                          & Wi-Fi  &  1.6    & 3.5  & After RB1 & 316  & 74.1\\
\hline
\end{tabular}
\end{table*}
Table~\ref{table:layer_split} presents the latency and mobile energy consumption by placing the bottleneck unit with obtained $c'$ and $s$ values (for the accuracy loss less than 2\%) after each residual block, for different wireless networks when there is no congestion in the mobile, cloud, and wireless network. The best partitions in terms of end-to-end latency and mobile energy consumption across different wireless settings are highlighted in this Table. Table~\ref{tab:lateny_energy_results} compares highlighted partitions with the mobile-only and cloud-only approaches in terms of latency and energy. For the cloud-only approach, before transmitting the input to the cloud, we apply JPEG compression on the input images which are stored in 8-bit RGB format.
Note that the best partitioning for the goal of minimum end-to-end latency is the same as the best partitioning for the goal of minimum mobile energy consumption in each wireless network settings. This is mainly due to the fact that end-to-end latency and mobile energy consumption are proportional to each other since the dominant portion of both of them are associated with the wireless transfer overheads of the intermediate feature tensor.

\textbf{Latency Improvement} - As demonstrated in Table~\ref{tab:lateny_energy_results}, using our proposed method, the end-to-end latency achieves 63$\times$, 21$\times$, 8$\times$ improvements over the cloud-only approach in 3G, 4G, and Wi-Fi networks, respectively.

\textbf{Energy Improvement} - As demonstrated in Table~\ref{tab:lateny_energy_results}, using our proposed method, the mobile energy consumption achieves 47$\times$, 41$\times$, and 31$\times$ improvements over the cloud-only approach in 3G, 4G, and Wi-Fi networks, respectively.

\textcolor{black}{As observed in Table~\ref{table:layer_split}, the best partition across all wireless network settings is associated with placing the bottleneck unit after RB1. This is an important result since as explained before and according to Fig. \ref{fig:resnet_feature_size}, due to the relatively large sizes of intermediate feature tensors in ResNet-50 compared to the input image size (especially RB1 which has the largest feature size), merely splitting the network between the mobile and cloud and transmitting the intermediate feature tensor to the cloud may not perform better than the cloud-only approach in terms of latency and mobile energy consumption. However, using our proposed method, mobile device can compute only one RB and send the reduced dense output feature tensor to the cloud, achieving minimum latency and mobile energy consumption among all possible partitions and significant improvements compared to the cloud-only approach, while the acceptable accuracy is still reached.}
\subsection{The Efficacy of Compression-aware Training} \label{section.efficacy_training}

\begin{figure}
\begin{center}
  \includegraphics[width=\columnwidth]{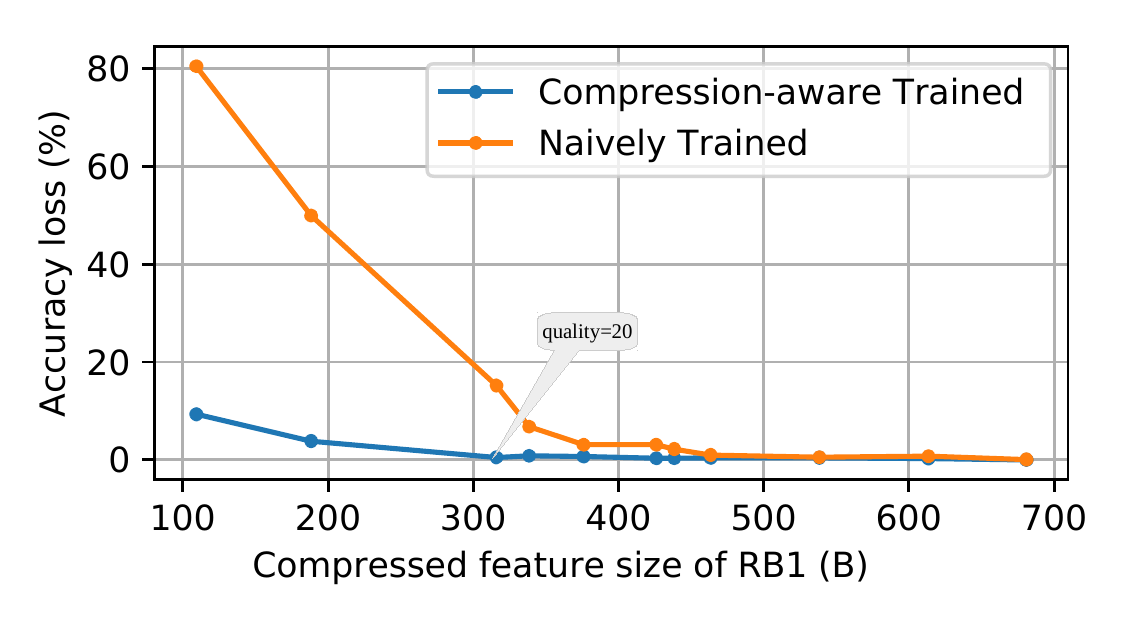}
  \caption{The comparison between the accuracy loss of the proposed compression-aware training and a naively trained model for different compressed feature size values, when the bottleneck unit is placed after RB1.}
  \label{fig:compression_aware_training}
\end{center}
\end{figure}

Incorporating the pair of JPEG compressor and decompressor as a new computational unit in a neural network can be performed in two ways: 1) Placing the compression unit in a given trained model (Naive), 2) Training the model from scratch using the proposed compression-aware training method as explained in Section~\ref{section.compression-aware_training}. 
\textcolor{black}{In JPEG compressor, which is the choice of lossy compressor for our experiments, changing a parameter named \textit{quality} affects the amount of output bits of the compressor. Fig.~\ref{fig:compression_aware_training} presents the accuracy loss obtained for ResNet-50 when the bottleneck unit is placed after RB1, versus different number of output bits of the compressor (corresponding to different values of the JPEG quality parameter, ranging from 1 to 100). }
As depicted in Fig.~\ref{fig:compression_aware_training}, the accuracy loss of the compression-aware training becomes almost zero by having a JPEG quality level of higher than 20, while the accuracy loss of the naive approach is close to 18\% using the quality level of 20. In our experiments, we use the quality level of 20 for JPEG compression in order to achieve maximum bit savings while there is no accuracy loss.

\subsection{Server Load Variations}
Data centers usually experience fluctuating load patterns. High server utilization can lead to increased service times for DNN queries. Therefore, it can be desired sometimes to change the point of partition at the run time and push more work toward the mobile device to decrease the current server load level. In order to allow for flexibility in the dynamic selection of partition points, both the mobile and cloud can host all possible $M$ partitioned models obtained via the training phase of algorithm \ref{the_algorithm}. For each of $M$ models, the mobile and cloud store only their assigned computing portion of the inference network. Depending on the server load, using the profiling and selection phase of algorithm \ref{the_algorithm}, partition point can be changed while still providing an acceptable accuracy and offloading less data compared to the cloud-only approach. This shows the efficacy of the learnable reductions in channel and spatial dimensions and using the compression-aware training method to avoid long latency of DNN queries caused by high user demands and server congestion. The best model (partition) can be selected at run-time by the mobile by periodically pinging the server during the mobile idle period. 



\subsection{Comparison to Other Feature Compression Techniques}
In comparison with other works in collaborative intelligence framework which have considered the compression of intermediate features before uploading them to the cloud, our proposed method can achieve significantly higher bit savings compared to the cloud-only approach. For instance, as reported in \cite{choi2018deep}, which is one of the few works in collaborative intelligence literature which consider lossy compression of features before transmitting them to cloud, communication overhead in their work can be reduced up to 70\% compared to the cloud-only approach.
However, in our work, with the proposed trainable reduction unit for spatial and channel dimensions, and the proposed lossy compression-aware training method, we can achieve up to 84$\times$ bit savings compared to the cloud-only approach according to Table \ref{tab:lateny_energy_results}. 
This shows that in collaborative intelligence framework, adding a learnable reduction in channel and spatial dimension alongside with a compression-aware training method can significantly perform better than merely splitting a network with fixed weights and compressing the intermediate feature tensor before uploading to the cloud.
\section{Conclusion and Future Work} \label{section.conclusion}
Recent studies have shown that the latency and energy consumption of deep neural networks in mobile applications can be considerably reduced by splitting the network between the mobile and cloud in a collaborative intelligence framework. In this work, we develop a new partitioning scheme that creates a bottleneck in a neural network using the proposed bottleneck unit, which considerably reduces the communication costs of feature transfer between the mobile and cloud. Our proposed method can adapt to any DNN architecture, hardware platform, wireless network settings, and mobile and server load levels, and selects the best partition point for the minimum end-to-end latency and/or mobile energy consumption at run-time. The new network architecture, including the introduced bottleneck unit after a selected layer of the underlying deep model, is trained end-to-end using our proposed compression-aware training method which allows significant bit savings while providing an acceptable accuracy. Our proposed method, across different wireless network settings, achieves on average 30$\times$ improvements for end-to-end latency and 40$\times$ improvements for mobile energy consumption compared to the cloud-only approach for ResNet-50, while the accuracy loss is less than 2\%.

For future work, the adaptability of our proposed method to various underlying deep models should be studied, as well as the usage of other compression techniques rather than JPEG in the bottleneck unit. Furthermore, the extent of reduction in the feature tensor dimension can be explored further using other architectures for the learnable reduction unit.


\bibliographystyle{unsrt}
\bibliography{references}

\end{document}